**MH3**

# Shifting Role of Customer from Recipient To Partner of Care In Healthcare Organization

Muhammad Anshari [1] and Mohammad Nabil Almunawar [2+]

[1] Universiti Brunei Darussalam & FST-UIN Yogyakarta Indonesia
[2] Universiti Brunei Darussalam

[1] 10h0104@ubd.edu.bn
[2] nabil.almunawar@ubd.edu.bn

**Abstract-** Most recent e-health initiatives perceive customers (patients) as recipients of medical care where they do not have a significant role in the process of health decision making. However, the advancement of Web 2.0 offers patients to have a greater role in the decision making process related to their health as they can be empowered with the ability to access and control information that fits with their personalized needs. However, providing patient empowerment in e-health through Web 2.0 is challenging task because the complexity nature of healthcare business processes. Empowerment closely relates to the concept of Customer Relationship Management (CRM) in managing good relationships with the customers. The adoption of Web 2.0 in CRM systems is known as Social CRM or CRM 2.0. Social CRM emerges to accommodate dynamic means of interaction between patients with their healthcare providers. The aim of this paper is to present a model that embeds empowerment of patient through Social CRM intervention that may extend the role of the patient as an individual health actor, a social health agent, and a medical care partner. A survey has been conducted to gain a feedback from customers regarding the proposed model. A prototype derived from the model namely Clinic 2.0 has also been developed. Using the prototype we measure its impact towards customer satisfaction and health literacy. The results show that the system intervention through Clinic 2.0 improves the level of satisfaction and health literacy of participants.

## 1. INTRODUCTION

Many healthcare organizations suffer from some inefficiencies and inequities in both service provision and quality (Garrrido& Acevedo, 2010). Some of these problems are due to the poor service management and the information flows (Kirsch, 2002). A successful healthcare organization exists when it provides good quality of service and keeps offering best quality services. In a competitive environment, the effort to attract a customer (patient) can take a month or even more but easy to lose one. So there must be values of pleasing a customer, satisfying the

[+] Muhammad Anshari is Lecturer at Dept. Informatics Engineering Faculty of Science & Technology UIN – Yogyakarta, Indonesia; Dr. Mohd.Nabil Almunawar is Senior Lecturer & Vice Dean at Faculty of Business, Economic, and Policy Studies Universiti Brunei Darussalam, Jl.Tungku Link Gadong, BE 1410 Brunei Darussalam. Tel.: + 673 8825711; fax: +673 2465017. *E-mail address*: anshari@yahoo.com, nabil.almunawar@ubd.edu.bn

customer's need, and build long-lasting relationship between customers and the organization that serves them (Low, 2002).

Adoption of Information and Communication Technology (ICT) in a healthcare organization can be aimed to improve the quality of service in healthcare information management (Chute et al., 1998). In addition, ICT drives the changes of healthcare's paradigm from 'Industrial Age Medicine to Information Age Healthcare' (Smith, 1997). This 'paradigm shift' is reshaping health systems and customers have been empowered in seeking information where the notions of healthcare services have been transformed from physical based services where they rely fully from physical presence to healthcare centre to home-based services where some healthcare activities can be performed at home, focusing on preventing diseases, promoting health, and give healthcare in their comfort times anywhere anytime (Haux et al., 2002).

With the advancement of Web technology and the wide adoption of Web 2.0 has brought a possibility to extend e-health service by accommodating social aspect such as enabling patients or patients' families, and the community at large to participate more actively in the process of health promotion and education through a social networking process. In fact, Web 2.0 has opened up opportunities to translate customers' empowerment in social media where they can share and discuss their concerns regarding health and healthcare.

An example of Web 2.0 application in healthcare is extending healthcare service through Social CRM or also called CRM 2.0 where many kinds of interaction can be supported such as interactions between a healthcare provider and its customers are among customers whereas customers are viewed as social health agents interacting through social media to share or discuss about health and healthcare issues that they experienced with. Web 2.0 encourages customers to take more responsibility for their own health, make decisions about their health concern, and help others through sharing. In turn, CRM initiatives with empowering features can be an effort to build loyalty and a trust between a healthcare organization and its customers.

In the traditional healthcare paradigm, customer or patient empowerment is neglected because a patient is viewed as a recipient of care while diseases and treatment of the patient are solely decided by his/her healthcare provider. This implies there is very minimal or no participation from the patient. This mindset is the main obstacle for the empowerment of patients. However, there will always be circumstances in which patients choose to hand over responsibility for decisions about their healthcare to providers. It is due to the difficulty of the choices, or the time involved in gaining an understanding of the health problem and the options, this does not undermine the proposition that a customer or patient empowerment will promote efficiency and that decisions should be made from the perspective of the customer (Segal, 1998).

Gibson (1991) defines empowerment is a process of helping people to assert control over the factors which affect their lives. It encompasses both the individual responsibility in healthcare and the broader institutional or societal responsibilities in enabling people to assume responsibility for their own health. McWilliam et al. (1997) view empowerment as the result of both an interactive and a personal process, where the emergence of ''power'' (or potential) is facilitated by a caring relationship, and not merely given by someone, nor created within someone. In other words, the emergence of a person's potential because of an empowerment process may view as a co-creation, within a true partnership (Low, 2002).

In terms of empowering individual in e-health service, Australia is a pioneer with the introduction of Personally Controlled Electronic Health Record (PCEHR). PCEHR is enabling patients in Australia to view their medical records online. However, so far there are not many literatures discussing the issue of empowerment that integrate individual, social, and medical aspects. Therefore, there is knowledge gap to address on how healthcare providers develop a mechanism(s) in encouraging customer's responsibility to take a greater role in decisions about their own care and delivery arrangements that will empower them in healthcare service delivery to meet increasing demands and expectations of customers while optimizing the cost of service. Recent discussion of empowerment is supported in the health literature, and it has been used for customers and healthcare services over the past decade. The proposed model is developed to enhance existing theory of empowerment in e-health business processes with the help of recent Web technology. The model integrates the wider scope of empowerment in healthcare (personal, social, and medical), value configuration of e-health's business process, electronic medical record, and adoption of Web 2.0. The model is expected to contribute in determining dimensions of e-health business process with the possible perspective of empowerment.

There are five stages to accomplish the study. Research design is started from analyzing previous studies on the related work in the literature. Then, we propose a reference model based on the literature analysis. A survey was derived from proposed model to capture user expectation. The result of survey is used as requirements in developing the prototype. Finally, the prototype is being used to in a real healthcare scenario testing in order to measure health literacy and customer satisfaction of participants.In the next section, we present a literature review of related work and the proposed model, Section 3 contains survey results and findings from the testing,and finally Section 4 is the conclusions.

## 2. LITERATURE REVIEW

A health system includes all activities and structures whose primary purpose is to influence health in its broadest sense (Arah et al., 2006). This notion is in keeping with the WHO's use of the term health system: 'all the activities whose primary purpose is to promote, restore or maintain health (WHO, 2000). In the traditional healthcare practices, patient is seen as one who receives care, and the recipient of any medical decision(s) made. However, there is a paradigm shift of the patient's role from one who receives care to one who actively participates in the healthcare process. This section discusses interrelated references that discuss and become the foundation for the proposed model. Those references are value configuration to give a comprehensive picture of healthcare business process, Electronic Medical Record (EMR) is used to exercise the role of medical records in relation to empowerment, and Social CRM or CRM 2.0 is CRM concepts and strategies that featuring Web 2.0 technologies in providing a mechanism of social networks' empowerment.

### 2.1 E-Health's Business Process and Value Creation

Understanding a business process helps healthcare organizations appropriately deliver the strategic plan in providing in e-health service. It is important to examine each business process as a layer of value to the customers (patients/patients' family). Customers place a value on these services according to quality of outcome, quality of service, and price. The value of each layer depends on how well they are performing. When a healthcare cannot achieve its strategic objectives, it needs to engineer their activities to fit business processes with the strategy (Michael, 1994). If the business processes do not fit the strategy, it will diminish the value. For example, the value of a health promotion is reduced by a delay respond of patient's query or poor communication skills. The value of e-consultation in e-health service will reduce by a late response to the person in charge.

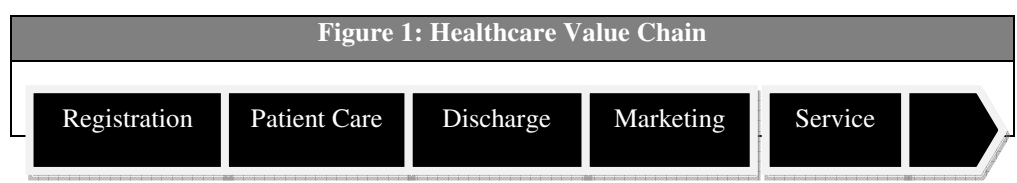

Figure 1: Healthcare Value Chain

Registration | Patient Care | Discharge | Marketing | Service

<div style="text-align: right;">Source: Porter, 1985</div>

Figure 1 above is originated from Porter's Value Chain analysis. Each activity adds value to the customer. These are registration, patient care, discharge, marketing, and service. Porter (1985) proposed value chain framework for the analysis of organizational level competitive strengths and weaknesses. It is a method for decomposing the firm into strategically important activities and understanding their impact on cost and value. According to Porter the overall value, creating logic of the value chain with its generic categories activities is valid in all industries. Later, Stabell and Fjeldstad (1998) proposed three distinct value configuration models that arguing the value chain is not valid in all industries. There are groups of industries that different in business nature and resource allocation.

According to Stabell and Fjeldstad(1998) refined the three distinct generic value configuration models (value chain, value shop, and value network) required to understand and analyze firm-level value creation logic across a broad range of industries. With the identification of alternative value creation technologies, value chain analysis is sharpened a value configuration analysis to gain competitive advantage. Stabell and Fjeldstad(1998) suggested that the value chain models the activities of long-linked technology, while the value shop models firm where value is created by mobilizing resources and activities to resolve a particular customer problem. Healthcare, schools & universities, and consulting firms are examples of value shop that rely on an intensive technology.

The medical consultation shop appears to be diagnosis-focused shop. Treatment plans follow the diagnostic. The cyclic nature of the activities set is captured by the circular layout of the primary activity categories. The five generic categories of primary value shop activities; Problem-finding and acquisition – activities associated with the recording, reviewing, and formulating of the problem to be solved and choosing the overall approach to solving the problem, problem finding – activities associated with generating and evaluating alternative solutions, choice – activities associated with choosing among alternative problem solutions, execution – activities associated with communicating, organizing, and implementing the chosen solution, and control & evaluation – activities associated with measuring and evaluating to what extent implementation has solved the initial problem statement (Michael, 1994). The value shop proposed by Stabell and Fjeldstad (1998) was adopted and modified accordingly in e-health scenarios to make individually (patient) is a partner in the care process. Set of activities is in looping process to show the interdependence between one activity with another and repetitive process of each individual.

## 2.2 Social CRM / CRM 2.0

Empowerment closely relates to the managing relationship between customers (patients) and healthcare provider. CRM is the concept and strategy used to manage those relationships. Looking into the value chain at Figure 1, CRM may exist from registration up to services at the end of activities. In conventional CRM, healthcare staffs almost do all activities from registration to services, and there is limited activities where customers (patients) are empowered to involve actively in each process. For example, the ability of customers to register online is kind of empowerment, therefore customers can be authorized to do by themselves for e-registration.In case of personal health activities, how is it possible for patients to record their daily habits by themselves if there is no systems provided that empower them to recordthose activities by their own without relying to medical staffs. Both scenarios are the examples on how CRM can provide a mechanism of empowerment in managing the relationship between patients – healthcare provider through e-health systems.

A next generation of CRM known as Social CRM or second generation of CRM (CRM 2.0) is a concept with the featuring Web 2.0 brings a promise to extend the idea of empowerment in e-health with social network capabilities.Web 2.0, which plays a significant part in the CRM transition, is a revolution on how people communicate. It facilitates peer-to-peer collaboration and easy access to real time communication and that is the core of social change (Almunawar et al., 2012b). Because much of the communication transition is organized around web based technologies, it is called Web 2.0 (Greenberg, 2009). The Web 2.0 is becoming a trend in Web technology. We are witnessing the acceptance of a second generation of web based communities such as wikis, blogs, and social networking sites which aim to facilitate creativity, collaboration, sharing among users rather than just for email and retrieve some information. Users can own the data on the Web 2.0 site and exercise control over that data (Hinchcliffe, 2006).

Greenberg (2009) defined Social CRM is a philosophy and a business strategy, supported by a technology platform, business rules, processes, and social characteristics, designed to engage the customer in a collaborative conversation in order to provide mutually beneficial value in a trusted and transparent business environment. It is the company's response to the customer's ownership of the conversation. While, Cipriani(2008) described the fundamental changes CRM 2.0 is introduced to the current, traditional CRM in terms of landscape. The most significant feature of CRM 2.0 is the network between customers and healthcare providers. This network creates value of a network such as multi-ways communications and sharing of experience and knowledge.

Table 1 summarizes the difference of CRM 2.0 from CRM 1.0 based on type of relationship, connection, and on how value generated. Relationship type in CRM 1.0 focuses on the individual relationship; Customer to Customer or Customer to Business whereas in CRM 2.0 offers the collaborative relationship and engage a more complex relationship network. Connection type in CRM 1.0 is a limited view of the customer that adversely affects the less informed customer; on the other hand, CRM 2.0 enables multiple connections that allow customers to be more understanding and knowledgeable. In terms of value creation, CRM 1.0 is constricted to targeted messages, whereas CRM 2.0 offers a more diverse value creation from informal conversation with customers within social networks.These are a few reasons why we consider CRM 2.0 for the empowerment process.

| Table 1: Comparison CRM 1.0 and CRM 2.0 | | |
|---|---|---|
| **Type** | **CRM 1.0** | **CRM 2.0** |
| **Relationship** | Focus on individual relationship (C2C , C2B) | Focus on the collaborative relationship (engaging a more complex relationship network) |
| **Connection** | Limited view of the customer & his community preferences, habits, etc. | Multiple connections allow better understanding of the customer and his community |
| **Generated Value** | Targeted messages generate value | Conversation generates value |

Source: Cipriani, 2008

CRM 2.0 is a collaborative conversation and sharing of their health experience with each other in order to provide mutually beneficial value. We refer customer in this activity as a social health agent where they engage in a collaborative relationship and conversation generates value for social supports.

### 2.3     The Proposed Model

The model constructed by modifying the process of value shop proposed by Stabell&Fjeldstad (1998) in healthcare services. The model accommodates dimension of empowerment in the form of personal, social, and medical. The cyclic model adopts from value shop's model that the process of healthcare is a repetitive and closed loop. Figure 2depicts the proposed model where the role of customers expands into three distinct functions as individual, social, and medical. Each role comprises of a set of sub modules that detailing the function and arranging activities within e-health's context between customers of healthcare provider. Because of the approach is modular, each stage is extendible depends on the need of healthcare providers. Therefore, from this modular approach, healthcare providers can vary in empowering their customers. Yellow circle in sub module indicates that provider empower their customers to have control on that specific sub module, while dash circle line indicates that provider only give partial empowerment to customers. In addition, no circle line means the healthcare providers provide no empowerment to the customers. From the perspective of the object oriented paradigm, the model composes of three districts object classes; personal, social, and medical (Figure 2). The impact of introducing object oriented approach in e-health systems can affect the process of medical or health recording. For instance, the conventional e-health system, medical records (EMR) are generated by healthcare staffs, but with the object-oriented approach, because the activities represent sets of objects then some objects' data can be empowered/delegated to customers (patients) in generating them. Here in the model, the terminology is introduced as Electronic Health Object (EHO). The term health is being used instead of medical simply because the scope of health is more comprehensive than medical. Therefore, EHO in this system comprises of object class personal, social, and medical.

      The object class personal derives function of *personal health actor*that exposes all personal health activities in e-health services as an individual (patient)which is directly or indirectly affect to their health status and services. Activities in this category are personal Identity (ID), personal Habits of patient (HB), Exercise activities (EX), Spiritual and Emotional activities (SE), personal Health Plan (HP), personal Account information (AC), and so on. For instance, ID sets of personal information in e-health system consists of personal information such as name, address, phone no, email address, login ID, password, etc. HB is the daily habit of individual that can include in this category such as eating, sleeping, and any other habits that may affect to personal health.  The EX is routinely exercising activities of individual that may benefit when they are recordedin the systems. All sub modules span in this category as discussed earlier can be empowered fully to the customers. They can manage by themselves. The process will replace a conventional approach instead of the health staffs normally will input patient information into their systems but the customers themselves will do it by their own. Obviously, it is empowering customers as a personal health actor for all activities under their own control, in other word, "give them the right what they can do it by themselves".

      The object class social empowers individuals and community as social health agentsfor others, which is considered as newly service in the e-health systems. Customer as a social health agent provides a broad range of empowerment by accommodating the concept of Social CRM/CRM 2.0) to utilize social networks within e-health services. Submodules in this category

are conversation (CS), chat, update status, forum, wikis, blog, knowledge management (KM), personal knowledge, group knowledge, asking for a specific service (RS). CS is standard social network activities such as sharing and conversation in social media. Nowadays, people use social networks in daily life. Updating status in social networks is triggering conversation among their circle of friends. Bringing this scenario into e-health services is an interesting issue and in fact challenging. For instance, patients with the same illness like diabetes may share their experiences with other patients in social networks. Sharing in social networks may become virtual supporting group that can enrich and strengthen their motivation to fight better health. In fact, social networks also affect to the relationship between healthcare provider and customers (patients).Therefore, adopting Social CRM in this category is imperative.

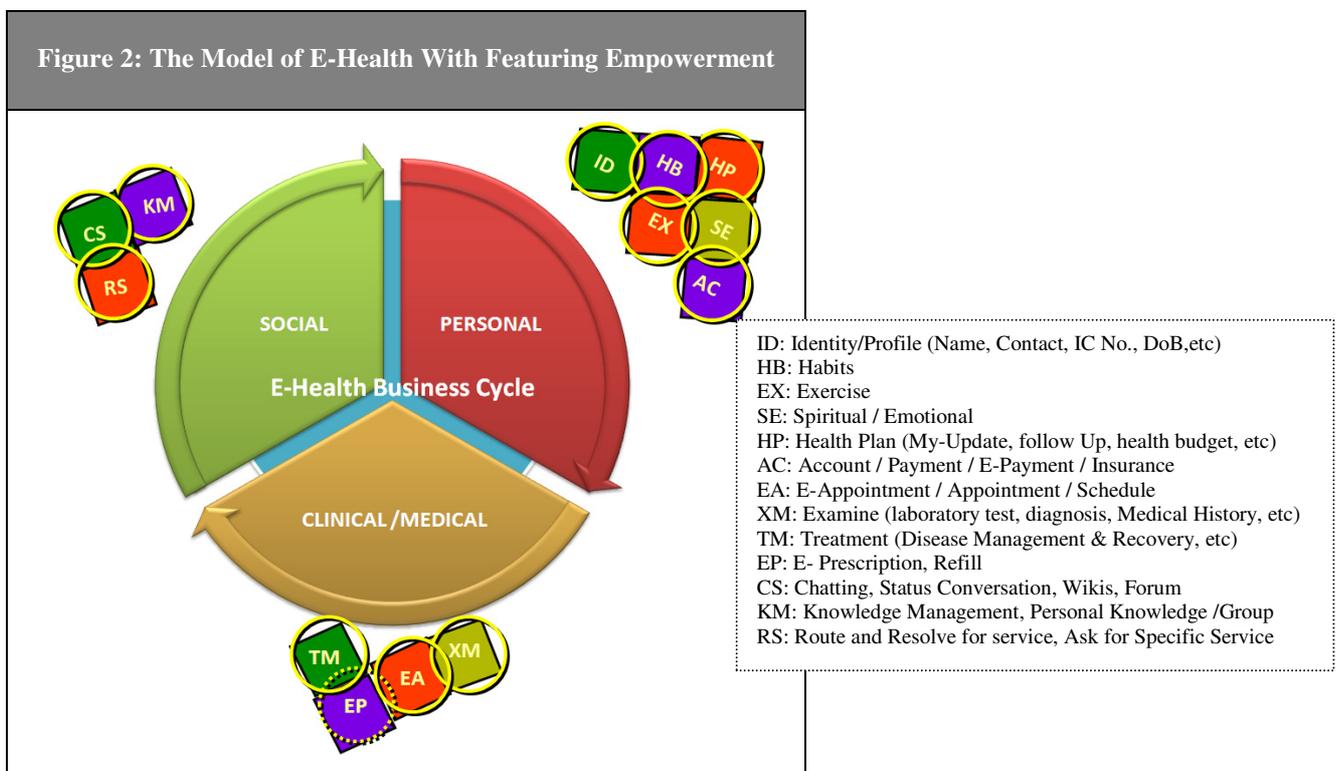

**Figure 2: The Model of E-Health With Featuring Empowerment**

ID: Identity/Profile (Name, Contact, IC No., DoB,etc)
HB: Habits
EX: Exercise
SE: Spiritual / Emotional
HP: Health Plan (My-Update, follow Up, health budget, etc)
AC: Account / Payment / E-Payment / Insurance
EA: E-Appointment / Appointment / Schedule
XM: Examine (laboratory test, diagnosis, Medical History, etc)
TM: Treatment (Disease Management & Recovery, etc)
EP: E- Prescription, Refill
CS: Chatting, Status Conversation, Wikis, Forum
KM: Knowledge Management, Personal Knowledge /Group
RS: Route and Resolve for service, Ask for Specific Service

Source (both picture and table): Authors' Compilation

Thirdly, object class medical is customer as *a medical health partner* in the process of healthcare services. This concept proposes the transformation of understanding and the role of patient in e-health business scenario especially the concept of empowerment. Submodules in this section are examined (XM), e-appointment (EA), e-prescription (EP), and e-treatment (TM). XM is an online consultation between patients and medical staffs that can lead to generate of electronic medical record (EMR) for the patients. It is a common service in any e-health initiative; however, when there is empowerment in these processes of medical activities, the result of e-consultation can be different with the e-health system without empowerment's features. For instance, many healthcare providers prevent patient to access their EMR prior to consultation time. So that those patients are not able to track their medical history by themselves, whenever the patient needs consultation, the diagnosis will start from beginning. In case of a patient, visit three different doctors for opinions, the patient will explain it all over again from beginning regarding their symptoms to those three doctors. E-health systems should educate patients regarding their health status, condition, and history. Therefore, empowerment in this category is main objectives to educate patients about their medical history and health status. The empowerment in this category may start from allowing customers to access their own EMR. The EP is another subset of EMR, which need also be empowered to customers. Empowerment in EP will speed up the process of managing prescription, and customers know how to consume due to they are able to access and learn anytime and anywhere. In summary, empowerment in online medical activities may shift the role of a patient from *recipient of care to partner of care*. Ability to access to knowledge and contents (EMR) make them partner for medical staffs, which is good for decision-making process for all.

In terms of ability for customers to generate content, each role and sub modules are able to producean electronic health record/contents that may help healthcare staffs in comprehensively diagnosis of a patient based on the generating contents by the customers. It is opposite with the conventional e-health system where the ability of customers (patients) to generate content that they perceive benefits unavailable to them. The model accommodates the type of empowerment into integrative interaction that is beneficial either for patient and healthcare organization. Integrating empowerment is an important features recognized as strategy for e-health services to improve health literacy and customers' satisfaction. Furthermore, the integrated approach can help healthcare organizations in defining which scopeof empowerment they will implement in the organization. While, modular approach will assist healthcare organizations to initiate empowerment by stages and later on to measure the empowerment process and performance modularly. The next section discusses the implementation of the model in the form of prototype of the system namely Clinic 2.0.

## 2.4 Clinic 2.0

The Clinic 2.0 describes the online tools and systems that facilitate interaction, exchange of information, and online contents made by web users. The system developed through the process of System Development Life Cycle (SDLC). SDLC is used in information systems, systems engineering, and software engineering as a process of creating new or altering existing systems. The SDLC can be thought of as a concept that lies beneath a number of software development methodologies currently employed throughout industry including healthcare information systems. The Clinic 2.0 is a prototype of e-health systems that implements the concept of Electronic Health Object (EHO) as discussed in previous section. It implements paradigm of an interactive healthcare system where any business process is represented as an object that has its own data, attribute, and method to act and interact with other objects. Though the system is not complete modules as proposed in business architecture (Figure 2), nevertheless it extends multi-ways of patients' interaction. Three possible relationships and interaction involve in the system are patient alone (patient with the systems interaction), patient to patient interactions, and patient with healthcare provider interactions.

The main menu is user friendly and clear navigation design allows users to find and access information effectively. At the top menu consists of a search box, My Health, Medical Record, Logout, Message of The Day (MOTD), status, conversation, profile, and group. The search gives visitors a quickly find the information they need. My Health is representation of object class personal in the business process design. It represents the cyclic model activities as shown in Figure 2. Medical Record is representation of object class medical or medical activities as shown in snapshot Figure 3.

The quick logout component adds a new menu type to Clinic 2.0 that allows a single click "logout" menu item (without requiring user confirmation to logout). One of the special feature in home menu is the MOTD, the MOTD is a message that is managed by health educator to send a reminder message to each user (patients or staffs) that is customized based on the need of each user. It means MOTD is uniquely different for each patient. For instance, the patient with diabetes will be likely to receive a note or reminder from healthcare educator based on his health condition. The MOTD is updated at regular time by the online health educator. Next after MOTD is notification for the user. The notification appears when there is friends request or new message to Inbox either from other users. The next section is status updates also known as a "status" allows users to post personal messages for their friends to read or share with others. In turn, friends can respond with their own comments, as well as clicking the "Like" button. A user's most recent updates appear at the top of their Timeline/Wall and is also noted in the "Recently Updated" section of a user's friend list. The purpose of the feature was to allow users to inform their friends of their current "status".

Right below after update status is a news feed, when users logged into Clinic 2.0, they are presented with a customizable version of their own profile. The new layout, by contrast, created an alternative home page in which users saw a constantly updated list of their friends' activity. News Feed highlights information that includes profile changes, upcoming events, and birthdays, among other updates. On the left side, there are menus of favorites, profile, group, suggestion friend, and online friend. Favorites consists of messages, knowledge, and forum. The message is menu to send and receive messages from other users. Knowledge is a resource center for all users where online health educators, medical staffs, and physicians share topics on health and medical treatment. Patients can only ask the topic being posted by the online health educators. While forum is the medium of exchange and sharing where patients can post and share any knowledge and experiences.

The different between menu knowledge and forum is on the quality of information being posted. In knowledge menu, the information about health and medical are reliable because its source from medical staffs or online health educator and the knowledge management is maintained to ensure the quality of information. While, the forum is unverified information because the source of patients that have not been tested in clinical evidence. The profile menu consists of basic information, profile picture, friends and family, education and employment, philosophy, entertainment, sport, interest and hobby, and contact info. They can also join interest groups. The group menu is optional for patients to join any available group of social networks such as diabetes, cancer, heart disease, etc. . Clinic 2.0 allows different networks and groups which many users can join. This is essentially equivalent to control of a blog for the administrators. Suggestion friend menu shows an option to invite others becoming a friend on social networks and group. Finally, the online friend menu shows friends who currently online. The user can initiate chat and conversation with online friends.

**Figure 3 Home Page Menu**

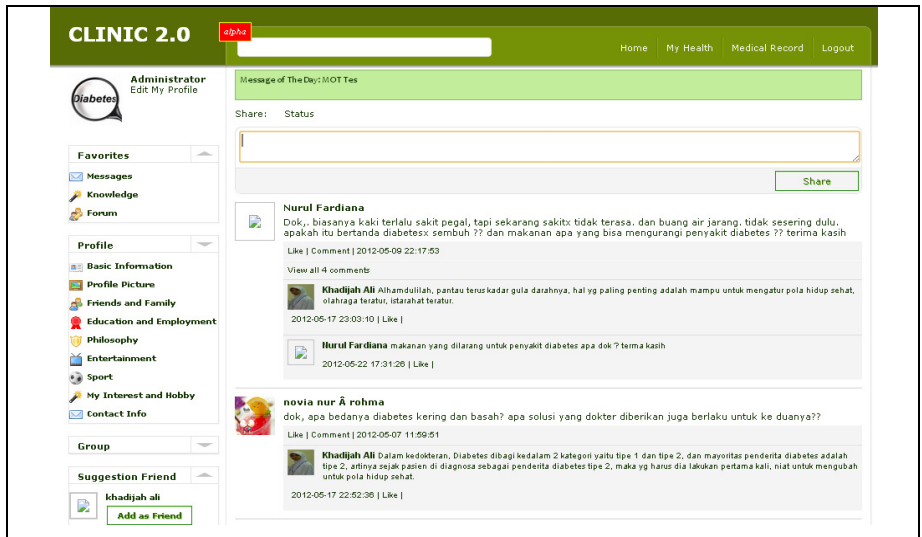

Source: Authors' compilation

## 3. SURVEY RESULTS AND FINDINGS

### 3.1 Background

Because no research has been published on the subject of e-health and empowerment in Brunei, this paper is prepared to fill that gap. Many empowerment topics in e-health are covered, including the use of existing e-health service, health information accessibility, and current issues like empowering through social support and social networking. First part was demographic information. Critical characteristics of participants such as their ages, genders, ethics and cultural identities, and socioeconomic situations, influence their health states, their access to healthcare, and the ways they are likely to use e-health innovations (Gustafson et al., 2001; Weiner et al., 2003). Table 2 is the demographic characteristics of the 366 respondents. Data gathered from the survey will be used to formulate recommendations for the future direction of empowerment in e-health systems.

| Table 2: Characteristics of Sample | | |
|---|---|---|
| **Item** | **Sample** | **Percentage** |
| **Employment** | Private | 42% |
| | Government | 58% |
| **Health Role** | Medical staffs | 21% |
| | Patients | 34% |
| | Patients' Family | 45% |
| **Gender** | Male | 46% |
| | Female | 54% |
| **Age** | 20 years or younger | 13% |
| | 21 - 30 | 38% |
| | 31 - 40 | 19% |
| | 41 - 50 | 18% |
| | 51 years or older | 12% |
| **Living Arrangement** | Alone | 90% |
| | Family | 5% |
| | Others | 5% |
| **Education** | Did not complete high school | 10% |
| | Completed high school only | 31% |

|  | Completed more than high school | 59% |
|---|---|---|
| **Internet Usage** | At least daily | 63% |
|  | <Daily to weekly | 18% |
|  | Weekly to monthly | 9% |
|  | Never | 10% |
| **Computer Usage** | At least daily | 64% |
|  | <Daily to weekly | 18% |
|  | Weekly to monthly | 9% |
|  | Never | 9% |

Source: Authors' Survey Sample

The first section of our survey questionnaire was designed to gather data on employment type, gender, age composition, living arrangements, level of education, frequency of computer usage, frequency of Internet usage. When it comes to the working pattern of the country, it is imperative to shed light on the respondents working in the government sector. This reflects the employment pattern in Brunei, where more people work with the government and the new generation is also inclined to work with the government for job security reasons and welfare benefits provided. Role of participants ranged from medical staffs, patients, and patients' family.It indicated the survey represented all potential users of the systems. In terms of educational level, each respondent has a different level of education. The difference in education level might reflect. This interprets that there are more educated respondents than those with lower education level. The results of the level of education might reflect the result of other factors such as their understanding towards the modern technology especially in ICT they are interrelated. Additionally, the respondents were questionedabout the frequency of their use of the Internet, regardless of what medium they use to get internet access. The results show that the majority of the respondents are exposed to the Internet and they tend to use the Internet every day.The respondents were also asked on how frequent they use the computer in their everyday lives for any reason(s). This shows that the computer literacy in Brunei is high and this statement supports the results for usage of Internet and educational level as well.

### 3.2 Survey Results

The model proposes empowerment of personal, social, and medical. The survey questioned was designed to gather the agreement and disagreement of participants if empowerment services are introduced in e-health systems. Table 3 summarizes the results of the survey that directly correlated with the empowerments. There are three sections clustering questions for empowerments.The question of medical empowerment is more than the other two because the core business of e-health is in medicine. For example, type of question that we have asked to the participant for medical empowerment, do you agree that you able to view your own electronic medical records (EMR) anywhere, anytime? Details of the agreement are shown in the percentage below.

| Table 3: Result of the Survey | | |
|---|---|---|
| **Survey Result** | **Sub Modules** | **Agreement in %** |
| **Personal Empowerment** | | |
| • Able to view health promotion online | HP | 81 |
| • Able to record personal health activities online | HB/EX | 76 |
| • Able to pay service online | AC | 50 |
| • Able to view payment online | AC | 73 |
| **Medical Empowerment** | | |
| • Able to view EMR | XM | 79 |
| • Able to request appointment online | EA | 83 |
| • Able to ask prescription online | EP | 77 |
| • Able to ask referral online | RS | 78 |
| • Able to consult online | XM | 73 |
| • Able to share EMR only for trusted doctors | XM/TM | 92 |
| **Social Empowerment** | | |
| • Able to discuss the health service in social networks | CS | 62 |
| • Able to control of own EMR | EA/XM/TM/EP | 77 |
| • Able to discuss the health status in a social network | CS | 72 |
| • Ability to discuss with patients same condition | KM/CS | 76 |

Source: Authors' Survey

There are four questions for personal empowerment. In fact, the majority of respondents show their agreement of personal empowerment features in e-health systems except for paying service online. It is reasonable because most of Bruneian enjoys with the facilities of free medical services in the country. When we asked about possibilities of recording their health related activities online like eating habits, 76% of respondents agreed to record their health-related activities online. It strongly indicates that they are willing to participate if the service is offered. These activities may include a *personal health diary* where the respondents can access it, anytime and anywhere, facilitating their daily plans and programs for a healthy lifestyle. This service can be used to monitor their own personal health habit, which may help them in making health decisions now or in the future. Therefore, respondents confirm that the majorityagrees to use the service that is being expectedto improve health outcomes of individual. Moreover, they approved to have access to these features because they will know better their own medical status, which may lead to be proactive patient, boost healthcare awareness, and enhance self-managed healthcare.

Interesting facts also depicted in the medical empowerment's data. More than two third of respondents responded positively of the medical empowerment in e-health. By looking into the demographic traits that the respondents are varies from medical staffs, patients, and patients' family. Interestingly, even medical staffs supported the ability of medical empowerment being introduced to customers. For instance, the ability to view EMR is the information onthe patient's medical history, laboratory test, diagnosis of previous conditions, etc. When we asked their opinions on the ability to view their medical history records online, the study shows that 79% agreed to view of their medical records online. This result indicates that they agree on the ability to self-monitor of their medical record history. In the comments, many respondents mentioned that the ability to access recorded medical history may lead to improve healthcare awareness and self managed healthcare.

The other example is consultation online, when patients are assigned to have a diet program, they do not have to travel all the way to a hospital because they can have access to the program using the online service provided and this saves time and money. For health professionals, this module will help them monitor their patients' progress online, which is convenient for both parties. Feature of electronic appointment (EA) EA is the ability of the patient to make an appointment online to book for medical visit online. The survey revealed that the majority of the respondents (83%) preferred to be able to make an appointment online. The result shows that the empowerment of MA's module is in line with their expectation(s) on the service. Also referred to EP's sub-module (E-Prescription, Refill), the majority of the respondents tend to agree on facilitates to request their preferred or trustworthy referral online. The majority of them agreed to view their payment and insurance covered online to keep up-to-date on their payment information and to avoid miscalculation. They agreed to communicate with their health promoter or health educator through ICT. It shows that most of Bruneians agreed on request refills for prescriptions online due to the time-efficiency it offers.

There are four questions for social empowermentwhere patients can share their health status after conducting their health program to their close relatives, which is also good to gain moral support or share with the trusted medical professionals.The survey confirms that customers agreed to discuss anything related to health services through social networks which they belong to because they want to share their experiences regarding on the variety of health services used with others. Sharing any issue relates to health service could improve customer service of healthcare organizations. Respondents are also willing to share their health condition with their friends in social networks, possibly because they want to know the point of view or advices of others regarding their health problems. Moreover, they may feel good when they can share their burden to other people who are experiencing the same condition. Having communicated with other patients who have the same condition will be beneficial to many people as they feel that they are not alone and they can discuss and share their experiences with others who have a similar problem. The social networking tools or online support group may become a medium to get or disseminate information as much as possible.

### 3.3 Testing on Health Literacy

This section was to produce an analytical result from the field study assessment. It is to analysis the results of twelve participants who were engaged in Clinic 2.0 testing process. The prototype of Clinic 2.0 was used to assess the level of health literacy and customer satisfaction of participants. It has been used by participants for about three months and then researcher compared the findings before and after interaction with the systems. Descriptive statistics were used for data analysis. Participants completed questionnaires before and after interacting with the Clinic 2.0. The comparison of customer satisfaction of participants was made through using the pre-test & post-test data. When the lowest score is 32 indicated as very unsatisfied in the highest scores is 81 as very satisfied.

### Health Literacy

The comparison of the health literacy of participants was made by using the pre-test & post-test data on CAHPS health literacy scoring. Based on CAHPS Health Literacy identifies the lowest score are 30 points indicated that the person has a very low level of health literacy while the highest level of literacy are 120 points (Figure 4).

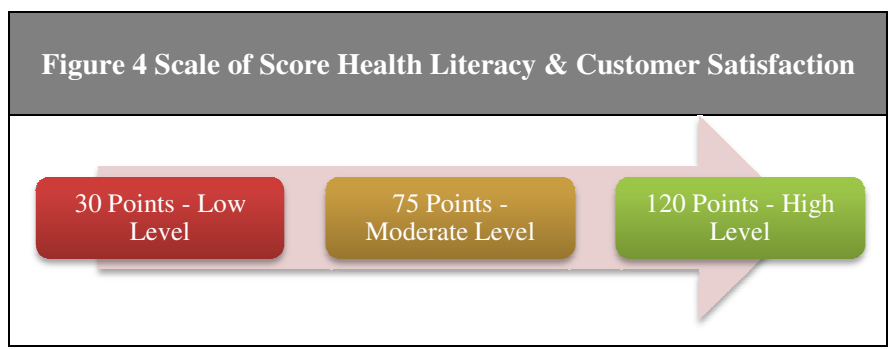

Figure 4 Scale of Score Health Literacy & Customer Satisfaction

PRE-TEST RESULTS

The score test of the participants of 30 questions of pre-test have shown that the mean of the health literacy of participants' were 75.25 (Table 4), and the Cronbach's Alpha is 0.96.

Table 4 Score of Health Literacy before Interaction

| Pre-test scores | Mean | Cronbach's Alpha | SD |
|---|---|---|---|
| 72, 76, 78,74,80,76, 74, 77,74,77,74,71 | 75.25 | 0.98 | 2.562 |

POST-TEST RESULTS

While, the mean of the participants after interaction with the system was 90.41 (Table 5). There was different between the score before and after the use of the Clinic 2.0. It was revealed that the interaction with the systems have improved their health literacy of participants.

Table 5 Score of Health Literacy after Interaction

| Post-test scores | Mean | Cronbach's Alpha | SD |
|---|---|---|---|
| 87, 90, 86,90,96,93, 91, 93,93,92,88,86 | 90.41 | 0.96 | 3.133 |

Figure 5 shows the difference of the survey result from twelve participants. The difference was not high as the participants were interacted with the systems for about three months only. We believe the longer they interactwith the system, the better improvement of their health literacy since in general (average) the participants have improved their health literacy after interract with the system (Clinic 2.0).

**Figure 5 Comparison the Score of Health Literacy for Pre and Post Interaction**

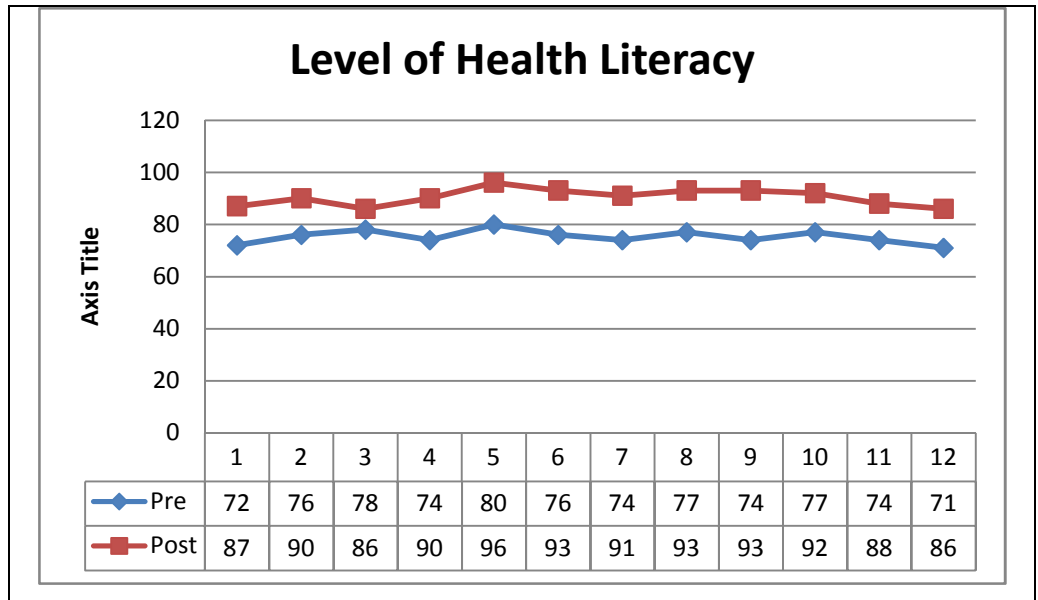

Source: Authors' Survey, 2012

### 3.4 Testing on Customer Satisfaction

The comparison of customer satisfaction of participants was made through using the pre-test & post-test data. When the lowest score is 32 indicated as very unsatisfied in the highest scores is 81 as very satisfied.

PRE-TEST SCORES

The score pretest of the participants has shown that the mean of the health literacy of participants' were 39.41 with the Cronbach's Alpha is 0.86 (Table 6).

Table 6 Score of Customer Satisfaction before Interaction

| Pre-test scores | Mean | Cronbach's Alpha | SD |
|---|---|---|---|
| 36, 52, 34,37,41,40, 39, 39,44,36,38,37 | 39.41 | 0.86 | 4.75 |

POST-TEST SCORES

The score post-test of the participants have shown that the mean of the health literacy of participants' were 71.33 with the Cronbach's Alpha is 0.97 (Table 7).

Table 7 Score of Customer Satisfaction after Interaction

| Post-test scores | Mean | Cronbach's Alpha | SD |
|---|---|---|---|
| 76, 75, 73,73,72,73, 71, 73,73,72,66,59 | 71.33 | 0.97 | 4.58 |

Figure 6 shows the difference of the survey result from twelve participants. The difference was very significant, it is because the comparison that the interactivity is never being facilitated before and now with the Clinic 2.0 is able to interact more closely with the system and healthcare provider. On average the participants improve significantly their level of satisfaction before interaction and after.

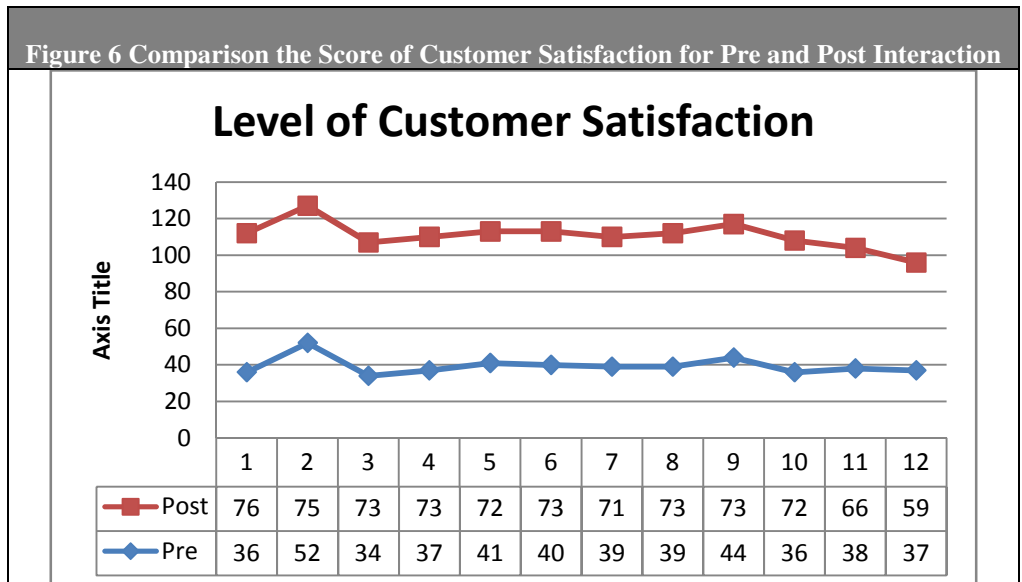

Source: Authors' Survey, 2012

## 4 CONCLUSION

E-health servicesin many ways and levels from the lowest level of digitizing medical record to the level where customers interactively can actively participate in full ranges of empowerment. Social CRM in e-health extends the role of customers (patients) into three dimensions of empowerment; individual health actor, social health agent, and partner in the medical care process. Simply because, it is able to managethe type of interaction between patient and healthcare provider may in the form of interaction between customers to provide like accessing online services, booking online consultation, paying online, etc. Patient-patient interactions are enables them to generate collaborative conversations in order to provide mutually beneficial value in a trusted environment. Each type of relationship entitles to empowerment capabilities in e-health systems. The model provides the ability to offer empowerment, ingenuousness/openness of information sharing, and closeness of the relationship between patient-provider and patient with others. The concept model generates value in each activity to the customer to provide better service. We believe with better service that generate values to the customer will create customers' trust and loyalty that helps customers and health providers in a sustaining relationship for mutual benefits. It is expected to contribute enhancing dimension of e-health business process with the possible perspective of empowerment.The survey and real testing scenario have validated that empowerment is in demand or in need. Their responds confirm and approve the list of empowerment abilities provided by the e-health system. The majority of respondents agreed with the three types of empowerment for individual, social, and medical in e-health setting. While, prototype testing confirmed that empowerment has contributed in their level of health literacy and customer satisfaction.